\documentclass[12pt]{book}

\usepackage[dvips]{graphicx,color}
\usepackage{makeidx,observe,cosmology}

\makeauthorindex
\makeindex

\BookTitle{New Trends in Theoretical and Observational Cosmology}
\CopyRight{\copyright 2001 by Universal Academy Press, Inc.}

\begin{document}

\BookTitle{\itshape New Trends in Theoretical and Observational Cosmology}
\CopyRight{\copyright 2001 by Universal Academy Press, Inc.}
\pagenumbering{arabic}

\chapter{
Primordial perturbations from inflation}

\author{%
David WANDS\\
{\it Relativity and Cosmology Group, School of Computer Science
and Mathematics, University of Portsmouth, Portsmouth PO1 2EG,
United Kingdom}}
%
%
\AuthorContents{D.\ Wands}
\AuthorIndex{Wands}{D.}

\section*{Abstract}

I review the standard analysis of adiabatic scalar and tensor
perturbations produced by slow-roll inflation driven by a single
scalar field, before going on to discuss recent work on the role of
non-adiabatic modes during and after inflation.  Isocurvature
perturbations correlated with adiabatic modes may be produced
during multi-field inflation and would give valuable information about
the physics of the early universe.  A significant contribution from
correlated isocurvature perturbations is not ruled out by present
data, but should be either detected or ruled out by future
observations.

\section{Introduction}

In recent years a standard model has emerged for the origin of the
large-scale structure of our Universe~\cite{LL}. Observations of the
cosmic microwave background (CMB) reveal primordial anisotropies on
the surface of last scattering of the CMB photons. Structure can form
from these initial perturbations about a Friedmann-Robertson-Walker
(FRW) universe by gravitational instability to form the galaxies,
clusters of galaxies and superclusters observed in large-scale
surveys.
These observations are consistent with an almost scale-invariant
initial spectrum of adiabatic density perturbations at
last-scattering~\cite{WTZ}.
Inflation is a dynamical model of the early universe which can explain
the origin of perturbations on arbitrarily large scales from
small-scale vacuum fluctuations of light fields.

Observational data are increasingly being used to constrain the
cosmological parameters of the background FRW model of the universe
since last-scattering. But this can only be done in the context of
some model for the nature of the primordial perturbations. Constraints
on the form of the primordial perturbations yield constraints on the
dynamics and high energy physics driving inflation in the very early
universe.

\section{Single-field inflation}

The simplest models of inflation are driven by a scalar field,
$\phi$, slowly rolling down its potential, $V(\phi)$, in a
spatially flat FRW spacetime with scale factor $a$. The classical
evolution is determined by the Klein-Gordon equation
\begin{equation}
\label{KGeqn}
 \ddot\phi + 3H\dot\phi = - V_\phi \,,
\end{equation}
(where $V_\phi$ denotes $dV/d\phi$)
coupled to the Friedmann equation for the Hubble expansion
($H\equiv \dot{a}/a$)
\begin{equation}
\label{Friedmann}
 H^2 = {8\pi \over3 M_P^2} \left( V + {1\over2}\dot\phi^2 \right) \,.
\end{equation}
This gives inflation (i.e., accelerated expansion, $\ddot{a}>0$)
for $V>\dot\phi^2$.

The slow-roll approximation truncates these to a first-order
system
\begin{eqnarray}
3H\dot\phi \simeq -V_\phi \,\\
H^2 \simeq {8\pi\over 3M_P^2} V \,.
\end{eqnarray}
This assumes the evolution is potential-dominated
($V\gg\dot\phi^2$) and over-damped ($3H|\dot\phi|\gg|\ddot\phi|$) but
can give a useful approximation to the growing mode solution when
the dimensionless slow-roll parameters~\cite{LL} are small
\begin{eqnarray}
\epsilon \equiv {M_P^2 \over 16\pi} \left( {V_{\phi}\over V}
\right)^2 \ll 1 \quad , \quad
|\eta| \equiv {M_P^2 \over 8\pi} \left| {V_{\phi\phi}\over V}
\right| \ll 1 \,.
\end{eqnarray}

Linear perturbations of a massless field in an FRW background obey
the wave equation
\begin{equation}
\ddot{\delta\phi} + 3H \dot{\delta\phi} - \nabla^2\delta\phi = 0
\,.
\end{equation}
Arbitrary inhomogeneities can be decomposed into spatial
harmonics, such as Fourier modes. Each mode with fixed comoving
wavenumber $k$ has two characteristic timescales
\begin{itemize}
\item oscillation period (determined by the physical wavelength) $a/k$
\item damping timescale (determined by the Hubble expansion)
$H^{-1}$
\end{itemize}
The evolution of each mode is thus naturally split into two
regimes
\begin{itemize}
\item
small scales ($a/k<H^{-1}$) under-damped {\it oscillations}
\item
large scales ($a/k<H^{-1}$) over-damped, or {\it frozen-in}
\end{itemize}
In a conventional (non-inflationary) matter- or
radiation-dominated universe the comoving Hubble length
$H^{-1}/a=\dot{a}^{-1}$ increases with time so that modes are
frozen-in ($k<aH$) at early times and only come within the Hubble
length at late times ($k>aH$). But in an inflationary era the
comoving Hubble length decreases and modes that begin as
under-damped oscillators on small scales are stretched by the
accelerated expansion beyond the Hubble length.
Thus initial zero-point fluctuations of the quantum vacuum on
small-scales (where the effects of the cosmological expansion is
negligible) leads to a spectrum of overdamped perturbations on
large-scales. Linear evolution ensures that the perturbations are
described by a Gaussian random field at all times.

Perturbations of a massive or self-interacting field have an
additional oscillation timescale set by the effective mass, $m^2\equiv
V_{\phi\phi}$. Massive fields ($m^2\geq9H^2/4$) remain underdamped
even on large-scales and effectively no perturbations are generated.
But any light field ($m^2<9H^2/4$) acquires a spectrum perturbations
$\langle\delta\phi^2\rangle\simeq (H/2\pi)^2$ at Hubble-crossing. In
particular, the inflaton must be light ($|\eta|\ll1$) during slow-roll
inflation.

\section{Cosmological perturbations}
\label{two}

Arbitrary perturbations of an FRW cosmology can be decomposed into
two types
\begin{itemize}
\item
{\it adiabatic perturbations} perturb the solution along the same
trajectory in phase-space as the background solution. Thus
perturbations in any scalar $x$ can be described by a unique
perturbation in expansion with respect to the background
\begin{equation}
\label{adiabatic}
 H\delta t = H{\delta x \over \dot{x}} \quad
\forall \quad x \,,
\end{equation}
e.g., the adiabatic density perturbation $\delta\rho/\rho\propto
H\delta\rho/\dot\rho$.
\item
{\it entropy perturbations} perturb the solution off the
background solution
\begin{equation}
\label{entropy}
 {\delta x \over \dot{x}} \neq {\delta y \over \dot{y}} \quad {\rm for\
 some}\ x\ {\rm and}\ y \,.
\end{equation} 
One example is an {\em isocuravture} perturbation of the baryon-photon
ratio $S=\delta(n_B/n_\gamma)=(\delta n_B/n_B)-(\delta
n_\gamma/n_\gamma)$.
\end{itemize}

Although the amplitude of adiabatic perturbations (such as the density
perturbation) is notoriously gauge-dependent, the adiabaticity
condition (\ref{adiabatic}) is not. Moreover, from this definition it
is clear that purely adiabatic perturbations (along the background
trajectory) must remain adiabatic on large scales and cannot generate
entropy perturbations (off that trajectory)~\cite{WMLL,GWBM}.

If the perturbed expansion, $H\delta t$, is evaluated in the
spatially-flat gauge~\cite{KodSas84} then it coincides with the
gauge-invariant scalar curvature perturbation, $\zeta$, on
uniform-density hypersurfaces~\cite{WMLL}, first introduced by
Bardeen, Steinhardt and Turner~\cite{BST83}. This is a
particularly useful quantity as it remains constant for adiabatic
perturbations in the large-scale limit.

In single-field inflation the over-damped perturbations of the
inflaton field on large scales are adiabatic perturbations of the
growing mode solution and entropy perturbations vanish in the
large-scale limit~\cite{GWBM}. Thus single-field inflation models
predict a primordial adiabatic perturbation on large-scales whose
amplitude can be calculated in terms of the scalar field
perturbations at Hubble-crossing ($k=aH$) during inflation. This
in turn can be related to the inflaton potential $V(\phi)$ in the
slow-roll approximation~\cite{Lidseyetal}
\begin{equation}
A_S^2
 = \left\langle \left( {H\delta\phi\over\dot\phi} \right)^2 \right\rangle_{k=aH}
  \simeq {32\over75}{V \over \epsilon M_P^4}
\end{equation}
The time-dependence of the inflaton field during inflation leads
to a scale dependence of the resulting spectrum, which in the
slow-roll approximation is determined by the slow-roll parameters
\begin{equation}
n_S - 1 \equiv {d\ln A_S^2 \over d\ln k} \simeq -6\epsilon + 2\eta
\,.
\end{equation}
In the extreme slow-roll limit this yields the scale-invariant
Harrison-Zel'dovich spectrum, $n_S=1$.

Gravitational waves, corresponding to tensor metric perturbations
independent of the scalar perturbations to linear
order~\cite{Bardeen80}, are another massless degree of freedom
excited during inflation and frozen-in on large scales to
give~\cite{Lidseyetal} 
\begin{equation}
A_T^2
 = \left\langle \left( {H \over M_P} \right)^2
 \right\rangle_{k=aH}
  \simeq {32\over75}{V \over M_P^4} \,,
\end{equation}
with spectral tilt
\begin{equation}
n_T \equiv {d\ln A_T^2 \over d\ln k} \simeq -2\epsilon \,.
\end{equation}
$A_S^2$ and $A_T^2$ describe the contribution of scalar and tensor
perturbations to the the CMB anisotropies on large angular scales.
A prediction of single-field slow-roll models of
inflation is that there should be a consistency condition relating
the scalar-tensor ratio to the tensor tilt~\cite{Lidseyetal}:
\begin{equation}
\label{consistency}
 {A_T^2 \over A_S^2} \simeq - {1\over2} n_T \,.
\end{equation}
Unfortunately there is no guarantee that the tensor contribution
is large enough to ever be detectable (let alone its tilt
measurable). Currently favoured hybrid-type inflation models generally
occur at low energies with $\epsilon\ll1$ and $A_T^2\ll
A_S^2$~\cite{LytRio}. 

\section{Non-adiabatic effects}

\subsection{Single field}

In single-field models the Hubble damping generally causes the
decaying mode solution to the Klein-Gordon equation to rapidly
decay, suppressing any non-adiabatic perturbations on large
scales.
Nonetheless, even in single-field inflation it is possible to
significantly alter the curvature perturbation on finite, but
super-Hubble scales. Non-adiabatic (decaying mode) and/or gradient
terms may have an effect if $z\equiv a\dot\phi/H$ (which is a
monotonic increasing function of time in the slow-roll
approximation) decreases back below its value at
Hubble-crossing~\cite{LSWL}. This is possible in some models of
inflation where the slope of the inflaton potential decreases
abruptly~\cite{Starobinsky} and the inflaton field enters a
transient friction-dominated `fast-roll' regime~\cite{LeaLid}
described by $\ddot\phi\simeq -3H\dot\phi$, leading to $z\propto
a^{-2}$ for a finite period.

Another case in which the instantaneous value of the scalar
curvature perturbation calculated at horizon-crossing may not be a
good estimate of the final value on large scales is when the
inflaton stops~\cite{SYK}, $\dot\phi=0$. In this case the apparent
divergence of $\zeta=H\delta\phi/\dot\phi$ is transient if at the
same time $V_{\phi}\neq0$ and, ironically, the slow-roll
approximation $\zeta\simeq V^{3/2}/V'$ can give a better estimate
of the final value, giving as it does an estimate of the
perturbation in the growing-mode solution~\cite{LSWL}.

\subsection{Two fields}

If more than one light scalar field exists during inflation
then there is more than one allowed phase-space trajectory for FRW
cosmologies. A spectrum of entropy perturbations (off the
background trajectory) will be generated on large scales from
initial vacuum fluctuations on small scales.

In the case of two canonical light fields $\varphi_1$ and
$\varphi_2$ the background trajectory is described by
$\dot\varphi_1$ and $\dot\varphi_2$ and arbitrary field
perturbations can be decomposed along and orthogonal to the
background trajectory~\cite{GWBM}:
\begin{equation}
\label{adent}
 \delta\sigma \equiv \cos\theta \delta\varphi_1 + \sin\theta \delta\varphi_2
 \,,\quad
 \delta s \equiv -\sin\theta \delta\varphi_1 + \cos\theta \delta\varphi_2
\end{equation}
where the angle of the trajectory in field-space is given by
$\tan\theta=\dot\varphi_2/\dot\varphi_1$. At any instant, the
adiabatic perturbation $\delta\sigma$ determines the scalar
curvature perturbation $R=H\delta\sigma/\dot\sigma$, while the
entropy perturbation determines the isocurvature perturbation
$S\propto\delta s$ at that time.

The coupled evolution equations for $\delta\sigma$ (in arbitrary
gauge) and $\delta s$ (which is automatically gauge invariant)
were derived in Ref.~\cite{GWBM} (for a different treatment of the
same equations see~\cite{Hwang}). In the slow-roll limit, and on
large scales, the equations can be reduced to~\cite{GWBM,BMR}
\begin{eqnarray}
\label{coupled}
 3H \dot{\delta\sigma} + \left( V_{\sigma\sigma} - \dot\theta^2
  \right) \delta\sigma &=& 2\left(\dot\theta\delta s\right)^. - 2
  \left({V_{\sigma}\over\dot\sigma} + {\dot{H}\over H}\right)
  \dot\theta\delta s \,,\nonumber \\
 3H \dot{\delta s} + \left( V_{ss} + 3\dot\theta^2
  \right) \delta s &=& 0
\end{eqnarray}
For $\dot\theta=0$ the equations are decoupled and reduce to the
standard slow-roll equations for canonical field perturbations.
But for a curved trajectory the entropy perturbation $\delta s$
appears as an extra driving term for $\delta\sigma$.
%
As a result the curvature perturbation is no longer constant on
large scales. This additional contribution to the spectrum of
scalar curvature perturbations at the end of inflation weakens the
consistency condition~(\ref{consistency}) of single-field
inflation to an inequality~\cite{PolSta,GBW,SasSte}
\begin{equation}
 \label{inconsistency}
 {A_T^2 \over A_S^2} \leq - 22 n_T \,.
\end{equation}


Langlois~\cite{Langlois} was the first to point out that another
consequence of this coupling is that any residual isocurvature
perturbation after multi-field inflation will in general be
correlated with the curvature perturbation.
The adiabatic and entropy field perturbations at horizon crossing
are, by their construction~(\ref{adent}), independent random
fields
\begin{equation}
 \langle \delta\sigma^2 \rangle_{k=aH} = \langle \delta s^2 \rangle_{k=aH}
 \simeq (H/2\pi)^2 \,, \quad
  \langle \delta\sigma \delta s \rangle_{k=aH} = 0 \,.
\end{equation}
The subsequent evolution on large scales can be parameterised by a
transfer matrix, which has the general form~\cite{Amendola}
\begin{equation}
\left( \begin{array}{c} R \\ S \end{array} \right)
 = \left( \begin{array}{cc} 1 & T_{RS} \\ 0 & T_{SS} \end{array}
 \right) \left(  \begin{array}{c} R \\ S \end{array}
 \right)_{k=aH}
\end{equation}
The form of this matrix is determined by the general definition of
adiabatic and entropy perturbations~\cite{WMLL} introduced in
section~\ref{two} and the evolution of two field perturbations
during inflation in Eq.~(\ref{coupled}) provides just one example.

The two coefficients $T_{RS}$ and $T_{SS}$ are model-dependent
transfer functions which determine the final power spectra
\begin{equation}
 \langle R^2 \rangle \propto 1 + T_{RS}^2 \,,\quad
 \langle S^2 \rangle \propto T_{SS}^2 \,, \quad
 \langle RS \rangle \propto T_{RS} \,.
\end{equation}
For example, if all the particle species present after inflation
are in thermal equilibrium determined by a single temperature then
there is a unique phase-space trajectory for the FRW cosmologies
and only adiabatic perturbations are possible, $T_{SS}=0$. The
other extreme is that one species remains completely decoupled
after Hubble-crossing corresponding to $T_{RS}=0$, leaving
uncorrelated isocurvature perturbations. But the general case is
that non-zero isocurvature perturbations survive and are
correlated with the curvature perturbations.

\section{Curvaton model for the origin of structure}

Taking into account the effect upon the large-scale curvature
perturbations of entropy perturbations possible in multi-field or
multi-fluid cosmological models leads to the realisation that the
curvature perturbation calculated at Hubble-crossing during inflation
only gives a {\em lower bound} for the curvature perturbation at the
start of the epoch of structure formation.
In a recent paper with David Lyth~\cite{LytWan}, I addressed the
question of whether it is possible to have a viable model of
structure formation if there is effectively no curvature
perturbation produced on large scales during inflation. The answer
is yes!

In our model the {\em curvaton} is supposed to be a light field during
inflation ($m_s^2\equiv V_{\,ss}\ll H^2$) which does not affect the
dynamics during inflation, i.e., is not the inflaton, and thus its
perturbations correspond to isocurvature perturbations. If this field
is decoupled from other inflation or matter fields then the
long-wavelength perturbations are effectively frozen-in until the
Hubble rate $H$ in the expanding universe drops to below the mass
$m_s$.  At this point the curvaton field begins to oscillate and its
energy density redshifts as $\rho_s\propto a^{-3}$. This grows
relative to the radiation density $\rho_\gamma\propto a^{-4}$ and will
come to dominate the energy density of the universe unless the field
decays.  This is the well-known Polonyi (or moduli) problem associated
with massive weakly-coupled fields. But a late-decaying field, so long
as it decays before nucleosynthesis, may not be a bad thing. Late
entropy release can dilute the abundance of other dangerous relics and
can be used as a model for baryogenesis or leptogenesis. Crucially,
its perturbations can also produce a large-scale curvature
perturbation~\cite{EnqSlo,LytWan,Moroi}.

As an example consider a complex scalar field
$\phi=|\Sigma|e^{is/v}$ whose modulus is fixed, $|\Sigma|\sim v$, by
a mexican-hat potential with large effective mass
$m_{|\Sigma|}\sim v$, but whose U(1) symmetry is only broken by
non-renormalisable terms so that $s$ has a small mass $m_s\sim
v^2/M_P$. The radial vev is stabilised, but the pseudo-Goldstone
boson $s$ will acquire an almost scale-invariant spectrum of
isocurvature fluctuations during a period of inflation, with Hubble
rate $H$, if we have
\begin{equation}
 {v^2\over M_P} \ll H \ll v \,.
\end{equation}
Assuming that after inflation $s$ decays only with gravitational strength,
$\Gamma\sim m_v^3/M_P^2$ then we naturally have
$\rho_s\sim\rho_\gamma$ at the decay time and hence~\cite{LytWan}
\begin{equation}
 \langle \zeta^2 \rangle \sim \langle (\delta s / s )^2 \rangle
 \sim (H/v)^2 \,.
\end{equation}

If the decay products thermalise completely then the
isocurvature perturbation during inflation, $\delta s$, is converted
into an adiabatic curvature perturbation, $\zeta$. It is then
indistinguishable from conventional inflaton models for structure
formation, other than violating the consistency
condition~(\ref{consistency}), but respecting the
inequality~(\ref{inconsistency}).

An interesting alternative possibility is that, because the curvaton decay
can occur relatively late, some particle species have already
dropped out of equilibrium when the curvaton decays. These species
would then have have an isocurvature perturbation relative to the
radiation produced by the curvaton decay, but one that would be
100\% correlated with the curvature perturbation.
A recent example of a late-decaying field that could produce
correlated curvature and isocurvature perturbations is the
sneutrino field in leptogenesis models~\cite{Yanagida}.

A similar model for the origin of large-scale structure from
initially isocurvature axion perturbations has been recently
proposed by Enqvist and Sloth \cite{EnqSlo} in the context of the
pre big bang scenario~\cite{LidWanCop}. In this case the rapid
increase in the Hubble rate during the pre big bang phase must be
compensated by the rapidly growing dilaton coupling to yield a
scale-invariant spectrum of axion perturbations~\cite{CEW}. This
appears to be the only possible origin of cosmological structure
in pre big bang type models where essentially no curvature
perturbation is produced on large scales during the `inflationary'
phase~\cite{LidWanCop}.

\section{Observational data}

The form of the primordial perturbation spectra is increasingly
being constrained by astronomical observations, especially cosmic
microwave background experiments. The overall amplitude of the
anisotropies on large scales is still set by the COBE
data~\cite{BLW96} which gives an amplitude
$
%
A^2 = 1.9 \times 10^{-5} \pm 10\%
%
$.
A recent complilation comparing data against FRW models with
adiabatic scalar and tensor perturbations by Wang, Tegmark and
Zaldarriage~\cite{WTZ} gives a constraint on the spectral index
$
 0.80 < n_S < 1.03
$,
and an upper limit on the contribution from gravitational waves,
$
 A_T^2 / A_S^2 < 0.008
$,
with the spectral index $n_T$ being unbounded.

Most studies of non-adiabatic perturbation spectra to date have
considered only uncorrelated isocurvature perturbations which tend to
only give an additional source of anisotropies on large angular scales
and hence their contribution to the CMB anisotropies is severely
constrained~\cite{EnqGB}. However I have emphasized that it is natural
in inflation models to consider isocurvature perturbations {\em
  correlated} with the standard adiabatic
mode~\cite{BMT,Durrer,Amendola}. In particular the cross-correlation
can decrease the CMB anisotropies on large angular scales and a
significant contribution from isocurvature modes cannot be ruled out
from current data. Indeed considering correlated CDM-isocurvature
modes (with scale-invariant correlation angle $\Delta$) the best-fit
to the current CMB data~\cite{Amendola} has $n_S=0.8$, $\cos\Delta=1$
and a similar contribution to large-angle anisotropies from adiabatic
and isocurvature modes. Future CMB data from the MAP satellite would
certainly be able to distinguish between such a model and a purely
adiabatic spectrum.

\section{Conclusions}

It is quite possible that the primordial perturbations observed by
future CMB experiments will remain consistent with the scale-invariant
Gaussian spectrum of adiabatic density perturbations proposed more
than thirty years ago by Harrison and Zel'dovich. In this case we
would be able to extract little about the physics of inflation and
would only be able to place bounds on allowed deviations from the
extreme slow-roll limit.

If we are to learn more about the dynamical history of the early
universe we need to find deviations from scale-invariance or traces of
either tensor perturbations or non-adiabatic effects which, though
not evident in current observations, could be detected by future
experiments.

\section*{Acknowledgements}

I am grateful to RESCEU at the University of Tokyo for their kind
hospitality. My work is supported by the Royal Society.



\end{document}